\documentclass[useAMS,usenatbib,usegraphicx]{mn2e}
\usepackage{amssymb}
\usepackage{latexsym}
\usepackage {keyval}
\usepackage{amsmath}

\title[Water maser outflow event in AFGL 2591 VLA~3-N]{Formation and evolution of the water maser outflow event in AFGL 2591 VLA~3-N}
\author[Trinidad et al.]{M. A. Trinidad$^{1}$\thanks{E-mail: trinidad@astro.ugto.mx (MAT); scuriel@astroscu.unam.mx (SC); robert@am.ub.es (RE); torrelles@ieec.cat (JMT); npatel@cfa.harvard.edu (NAP); jfg@iaa.es (JFG); guillem@iaa.es (GA); carrasco@mpifr-bonn.mpg.de (CC-G); raga@nucleares.unam.mx (ACR); l.rodriguez@crya.unam.mx (LFR)},
S. Curiel$^{2}$, R. Estalella$^{3}$, J. Cant\'o$^2$, A. Raga$^4$, 
\newauthor J. M. Torrelles$^5$,  N. A. Patel$^6$, J. F. G\'omez$^7$, G. Anglada$^7$, 
\newauthor C. Carrasco-Gonz\'alez$^8$, L. F. Rodr\'{\i}guez$^9$
\\
$^{1}$Departamento de Astronom\'{\i}a, Universidad de Guanajuato, Apdo. Postal 144, 36000 Guanajuato, M\'exico\\
$^{2}$Instituto de Astronom\'{\i}a (UNAM), Apartado 70-264, 04510 M\'exico
D. F., M\'exico\\
$^{3}$Departament d'Astronomia i Meteorologia and Institut de Ci\`{e}ncies del Cosmos (IEEC-UB), Universitat de Barcelona,\\~~Mart\'{\i} i Franqu\`{e}s 1, 08028 Barcelona, Spain\\
$^{4}$Instituto de Ciencias Nucleares (UNAM), Apartado 70-543, 04510 M\'exico D. F., M\'exico  \\
$^{5}$Instituto de Ciencias del Espacio (CSIC)-UB/IEEC, Universitat de Barcelona, Mart\'{\i} i Franqu\`{e}s 1, 08028 Barcelona, Spain\\
$^{6}$Harvard-Smithsonian Center for Astrophysics, 60 Garden Street, Cambridge, MA 02138, USA\\
$^{7}$Instituto de Astrof\'{\i}sica de Andaluc\'{\i}a (CSIC), Apartado 3004, 18080 Granada, Spain\\
$^8$Max-Planck-Institut f\"ur Radioastronomie (MPIfR), Auf dem H\"ugel 69, 53121 Bonn, Germany\\
$^{9}$Centro de Radioastronom\'{\i}a y Astrof\'{\i}sica (UNAM), 58089 Morelia, M\'exico}

\begin{document}

\date{Accepted 2012 December 21. Received 2012 December 20; in original form 2012 November 5}

\pagerange{\pageref{firstpage}--\pageref{lastpage}} \pubyear{2012}

\maketitle

\label{firstpage}

\begin{abstract}

In this paper we analyze multi-epoch Very Long Baseline Interferometry (VLBI) 
water maser observations carried out with the Very Long Baseline Array (VLBA) 
toward the high-mass star-forming region AFGL~2591. We detected maser
emission associated with the radio continuum sources VLA~2 and VLA~3.
In addition, a water maser cluster, VLA~3-N, was detected $\sim 0.5''$ 
north of VLA~3. We concentrate the discussion of this paper on the spatio-kinematical 
distribution of the water masers towards VLA~3-N. The water maser emission toward
the region VLA~3-N shows two bow shock-like structures, Northern and Southern,
separated from each other by $\sim 100$~mas ($\sim 330$~AU). 
The spatial distribution and kinematics of the water masers in this
cluster have persisted over a time span of seven years. The Northern bow shock has
a somewhat irregular morphology, while the Southern one has a remarkably smooth morphology.
We measured the proper motions of 33 water maser features, which have an average
proper motion velocity of $\sim$ 1.3~mas~yr$^{-1}$ ($\sim$ 20~km~s$^{-1}$).
The morphology and the proper motions of this cluster of water
masers show systematic expanding motions that could imply one or two different
centers of star formation activity. We made a detailed model for the Southern
structure, proposing two different kinematic models
to explain the 3-dimensional spatio-kinematical distribution of the water masers:
(1) a static central source driving the two bow-shock structures; (2) two independent
driving sources, one of them exciting the Northern bow-shock structure, and the
other one, a young runaway star moving in the local molecular medium exciting and
molding the remarkably smoother Southern bow-shock structure. Future observations
will be necessary to discriminate between the two scenarios, in particular by identifying the still unseen driving source(s).

\end{abstract}


\section{Introduction}

In the last few decades great progress has been made in the understanding of how low-mass stars form, through multi-wavelength observations and detailed theoretical models. It is now relatively well understood  that during the early stages of evolution of low-mass stars, a system is formed with a central protostar, surrounded by a rotating accretion disk at scales of $\sim$ 100 AU, and ejecting a highly collimated jet along the polar axis of the disk. 
While the disk is the reservoir from which the protostar accretes further matter, the ejected outflow removes angular momentum and magnetic flux from the system, allowing the accretion to proceed until the central star is assembled (e.g., Lada 1995; McKee \& Ostriker 2007; Machida, Inutsuka \& Matsumoto 2008; Armitage 2011; Williams \& Cieza 2011). However, the processes that give rise to massive stars ($\gtrsim$ 8~M$_{\odot}$) are still not well understood. Several well differentiated mechanisms for forming massive stars have been proposed. For example, via an accretion disk (as low-mass stars, but with a higher accretion rate),  via competitive accretion in a protocluster environment, or via mergers in very dense systems of lower mass stars (Bally \& Zinnecker, 2005).
Distinguishing between these possible scenarios is a very difficult observational task, essentially because massive young stars are rare and more distant compared to low-mass young stars. Furthermore, massive stars form in clusters in highly obscured regions, making it difficult to identify individual massive young stellar objects (YSOs) for detailed studies (see, e.g., Hoare \& Franco 2007; Hoare et al. 2007; Zinnecker \& Yorke 2007 and references therein).
In this sense, recent interferometer observations at cm and (sub)mm wavelengths
with angular resolutions of $\sim$ 0.1$''$ ($\sim$ 100 AU at 1 kpc distance) suggest that
stars with masses at least up to $\sim$ 20~M$_{\odot}$ may form via the accretion disk scenario, in a similar way as low-mass stars do, although the number of massive disk-protostar-jet systems identified and studied at scales of $\sim$ 1000 AU is still very small 
(e.g., Patel et al. 2005; Jim\'enez-Serra
et al. 2007; Torrelles et al. 2007;  Zapata et al. 2009; Davies et
al. 2010; Carrasco-Gonz\'alez et al. 2010, 2012; Fern\'andez-L\'opez
et al. 2011).

Within this context, observations of maser transitions of several molecular species (e.g.,
H$_2$O, OH, CH$_3$OH) in the proximity of high-mass protostellar objects
(HMPOs) provide a powerful diagnostic tool to investigate the first stages of the evolution of massive star formation (e.g., Sanna et al. 2010a,b; Bartkiewicz \& van Langevelde 2012). In particular, the use of Very Long Baseline Interferometry (VLBI) techniques for observing maser emission with angular resolution  $\la 1$~mas ($\la 1$~AU at 1~kpc distance), allows us
to study some of the main properties of the densest and hence most obscured portions of molecular clouds where the new massive stars are born. In addition, it is possible to derive the three-dimensional (3D) velocity distribution of the masing gas very close to HMPOs (e.g., Goddi et al. 2006; Matthews et al. 2010; Goddi, Moscadelli \& Sanna 2011; Torrelles et al. 2011; Chibueze et al. 2012), and through polarization measurements, the distribution and strength of the magnetic field 
in the cores of star forming regions (e.g., Vlemmings et al. 2006, 2010; Surcis et al. 2011a,b, 2012). Moreover, the detection of outflow activity by means of 
VLBI maser observations also allows us to identify new, previously unseen centers of massive star formation, some of them associated with 
unexpected phenomena such as ``short-lived" episodic ejection events characterized by kinematic ages of a few tens of years in the earliest stages of evolution of HMPOs (Torrelles et al. 2001, 2003, 2011; Gallimore et al. 2003; Surcis et al. 2011a; Chibueze et al. 2012; Sanna et al. 2012).

AFGL~2591 is one of the most extensively  studied high-mass star-forming regions in our Galaxy (e.g., van der Tak et al. 2006; Jim\'enez-Serra et al. 2012 and references therein). Located in the Cygnus X region, its distance has been recently estimated as $3.33\pm0.11$~kpc, via measurements of trigonometric parallax of masers (Rygl et al. 2012).
It appears completely obscured at optical wavelengths, 
but it has a very high luminosity in the infrared (IR) ($L_{\rm bol}\simeq2\times10^5~L_\odot$; Sanna et al. 2012). Three radio continuum sources have been detected toward AFGL~2591 within an area of $\sim$ 6$''$$\times$6$''$ ($\sim$ 0.1~pc), VLA~1, VLA~2, and VLA~3 (Campbell 1984; Trinidad et al.\ 2003).
Of these, VLA~3 is believed to be the youngest 
($\sim$ 2$\times$10$^4$ yr; Tofani et al. 1995; Doty et al. 2002; St\"auber et al. 2005) and more massive object (mass in the range 20-38~M$_{\odot}$), dominating the IR emission of the region (Sanna et al. 2012).  Submillimeter Array (SMA) observations carried out by 
Jim\'enez-Serra et al. (2012) show that VLA~3 is surrounded by a hot molecular core
 and that the global kinematics of the molecular gas is consistent with Keplerian-like rotation around a central source of $\sim$ 40~M$_{\odot}$. In addition, a powerful molecular outflow aligned along the east-west direction has been observed in
the region (Bally \& Lada 1983; Torrelles et al.\ 1983;  Lada et al.\ 1984;
Mitchell, Maillard \& Hasegawa 1991), with VLA~3 being its most likely  driving source candidate  (Trinidad et al.\ 2003). On the other hand, VLA~1 and VLA~2 are optically thin HII regions excited by early B-type stars (Trinidad et al. 2003).

\begin{figure*}
 \centering
 \includegraphics[width=168mm]{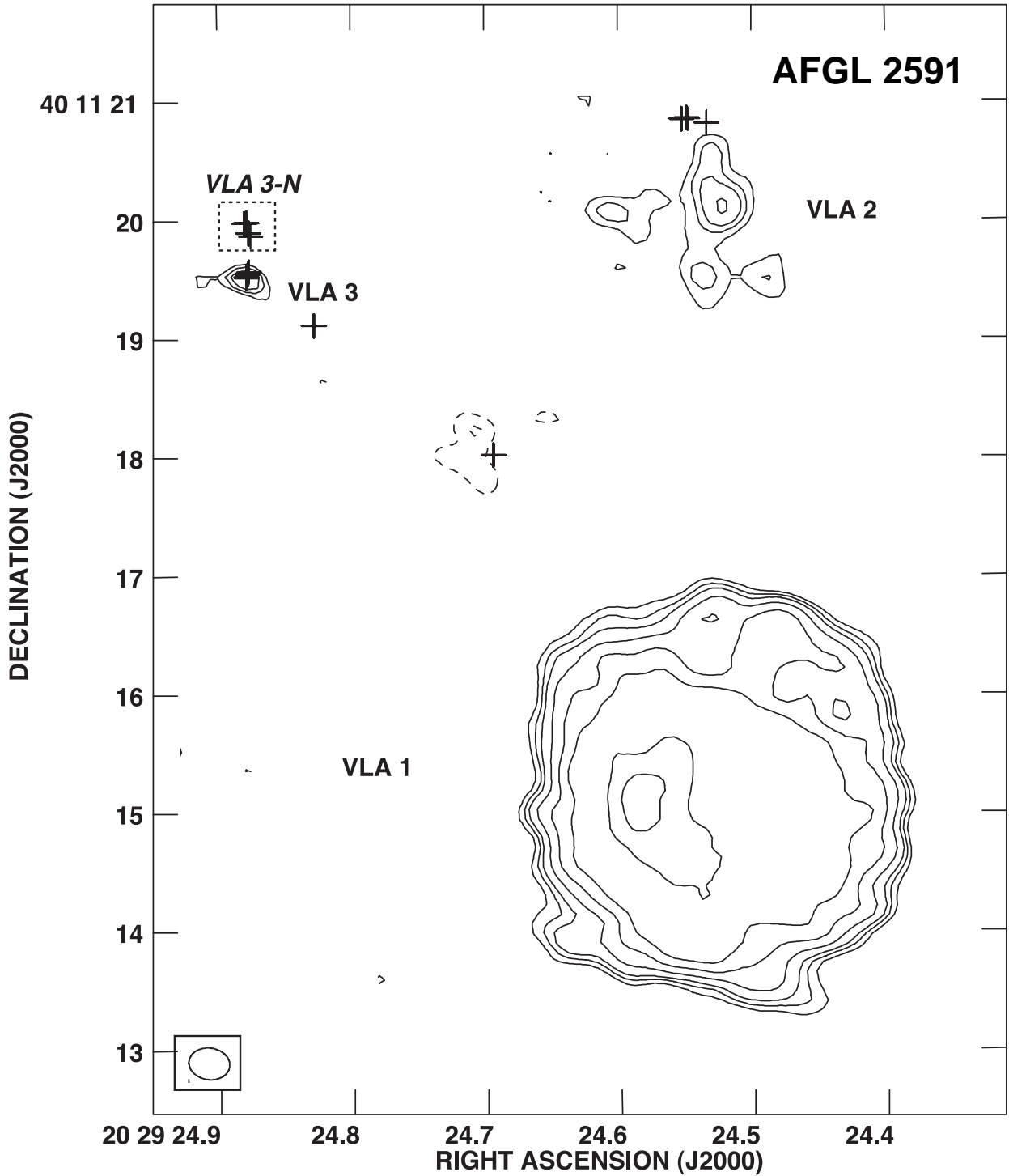} 
 \caption{VLA 3.6~cm continuum contour map of the high-mass star forming region AFGL 2591, showing the radio continuum sources VLA~1, VLA~2, and VLA~3, all of them 
excited by massive stars (see \S 1). Plus symbols  indicate the positions of the 
VLA 22~GHz water masers detected in the region (figure from Trinidad et al. 2003). 
The region analyzed and discussed in this paper is VLA~3-N (enclosed by a dashed square),
located $\sim$ 0.5$''$ north of VLA~3. This is the region where Trinidad et al. (2003) 
proposed that an internal source other than VLA~3 is exciting these water masers.}
\end{figure*}

Water maser emission is observed toward this massive star forming region, mainly associated with sources VLA~2 and VLA~3 (Tofani et al. 1995; Trinidad et al. 2003).  Trinidad et al. (2003)
detected with the Very Large Array (VLA; beam size $\simeq 0.08''$)
two well-differentiated clusters of water masers in the vicinity of VLA~3, one peaking at (and probably excited by) this massive YSO, and the other located $\sim$ 0.5$''$ ($\sim$ 1600 AU) north from it (see Figure 1). 
These authors suggest that the latter cluster is excited by an 
independent, still undetected embedded YSO located $\sim$ 0.5$''$ north from VLA~3. This interpretation has been recently supported by Sanna et al. (2012) with VLBI multi-epoch water maser observations (angular resolution of 0.6~mas), which show
 that the cluster of masers to the north of VLA~3 is formed by two bow shock-like structures separated $\sim$ 0.12$''$ ($\sim$ 400~AU), and moving away from each other along a north-south direction with proper motions of  $\sim$ 20~km~s$^{-1}$. Sanna et al. propose that a still undetected central massive protostar(s) (probably a late B-type star, based on the luminosity 
 of the masers) is ejecting a jet
that generates the two bow shocks at its tips.
The short kinematic age of the expanding motions traced by the 
water masers (few tens of years) also led these authors to suggest that they are probably the result of recurrent pulsed jet events rather than of a steady jet from the central protostar(s). 
Therefore, this could represent an important new case providing indications that massive objects are characterized by recurrent outflow events during the early stages of their evolution
(e.g. Mart\'{\i} et al 1995; Curiel et al. 2006).

In this paper we present (\S 2) VLBI multi-epoch water maser data  observed toward
AFGL~2591, about 7 years before (2001-2002 epochs) than those observed by Sanna et al (2012) (2008-2009 epochs). Our data show a water maser spatio-kinematical distribution that is globally consistent with that reported from the 2008-2009 epoch data, but  less extended and smoother. We analyze the 3-D kinematical and spatial distribution of the maser structures following their evolution during the time span of 7 years (\S 3), modeling the formation and evolution of these maser structures in terms of two possible scenarios: (1) a single static central massive YSO, ejecting a jet driving the two bow-shock  maser structures;  (2) two embedded driving sources, one exciting the northern bow shock, and the other one a young runaway star  moving in the local medium at $\sim$ 20~km~s$^{-1}$ and exciting the smoother southern bow-shock structure observed in our data (\S 4). The main conclusions are presented in \S 5.

\section{Observations}

Multi-epoch water maser observations were carried out with the Very Long Baseline Array
(VLBA) of the National Radio Astronomy Observatory (NRAO)\footnote{The NRAO is a facility
of the National Science Foundation operated under  cooperative agreement by Associated
Universities, Inc.} toward AFGL~2591 at three epochs (2001 December 2, 2002 February 11,
and 2002 March 5). We observed the $6_{16} - 5_{23}$ water maser transition (rest 
frequency = 22235.08~MHz) with a bandwidth of 8~MHz sampled over 512 channels (spectral
resolution 15.625~kHz = 0.21~km s$^{-1}$) and centered at $V_{\mathrm LSR} = -7.6$~km~s$^{-1}$.
All ten VLBA stations of the array were available
and recorded data during an observing time of 6 hours per epoch. The data were
correlated at the NRAO Array Operation Center. 

The amplitude and phase calibration of the observed visibilities, as well as further
imaging, were made with the NRAO Astronomical Image Processing System (AIPS) package. 
The sources 3C45, B2007+777, BL~Lac, and 3C454.3 were used for delay and phase calibration, 
while bandpass corrections were made using 3C345, BL Lac, and 3C454.3. 
An isolated maser spot, with a point-like morphology and with high intensity 
($\simeq$ 10~Jy) in all three epochs, was chosen to self-calibrate the data and to obtain 
a first and preliminary coordinate alignment between the three observed epochs. 
This reference maser spot has a radial velocity of $V_\mathrm{LSR}=-18.8$~km s$^{-1}$
and it is located toward VLA~3, with absolute coordinates 
$\alpha$(J2000.0) = {\rm 20$^h$29$^m$24.879$^s$}, $\delta$(J2000.0)
= 40$^{\circ}$11$'$19.47$''$ ($\pm$ 0.01$''$).

As a first step to identify the subregions of AFGL~2591 with water maser 
emission, we produced large maps with low angular resolution  for each of the three 
observed epochs. These maps had $8192\times 8192$ pixels of 1~mas each, over 512 velocity 
channels, covering a region of 8.2$''$$\times$8.2$''$ centered on the reference maser spot. 
From them, three subregions with maser emission were identified, one associated with VLA~2, 
another with VLA~3, and the third one located $\sim$~0.5$''$ north of VLA~3. 
We did not detect maser emission in the region $\sim 2''$ southwest of VLA~3, where water maser
emission was previously identified with the VLA (Trinidad et al. 2003; see Figure 1).
We then obtained simultaneously, for each epoch, full-resolution maps ($8192\times 8192$ 
pixels of 0.1~mas
each, over 512 channels) of these fields, as well as of an additional subregion 
where Sanna et al. (2012) also identified water maser emission (the subregion named in that paper as MW, and located $\sim$ 0.6$''$ west of VLA~3). The resulting synthesized beam size was $\sim$ 0.45 mas for the three observed epochs.
The rms noise level of the maps ranges from $\sim5$~mJy~beam$^{-1}$
in the channel maps with weak signal, to $\sim250$~mJy~beam$^{-1}$, in the
channels with the strongest maser components.

Finally, each velocity channel map was searched for maser spots. We refer  to a maser spot 
as emission that occurs  at a given velocity channel and with a distinct spatial
position. The threshold for the maser spot detection was taken to be equal to 10 $\sigma$
in each velocity channel, using the maser spot search procedure described in Surcis et al.
(2011a). Clusters of maser spots were detected in the subregions around VLA~2, VLA~3, and 
$\sim$ 0.5$''$ north of VLA~3.  We did not detect any emission in the MW region, where
 Sanna et al. (2012) found water maser emission  about seven years later. We fitted
all the maser spots detected with two-dimensional elliptical Gaussians,
determining their position, flux density, and radial velocity. 
The water maser emission in the AFGL~2591 region spans a velocity range from
$V_\mathrm{LSR} \simeq -31$ to $-2$~km~s$^{-1}$.  We estimated that the 1$\sigma$ accuracy
in the relative positions of the maser spots at each epoch is better than $\sim$ 0.01~mas
(beam/[2$\times$SNR]; Meehan et al. 1998). In this paper, we will concentrate on the spatio-kinematical 
distribution of the masers found $\sim$ 0.5$''$ north of VLA~3 (hereafter VLA~3-N), 
where Sanna et al. (2012) identified two bow-shock like structures (see \S 1). The other
two observed clusters of masers associated with VLA~2 and VLA~3 will be presented and 
discussed elsewhere.

\begin{figure*}
\centering
 \includegraphics[width=168mm]{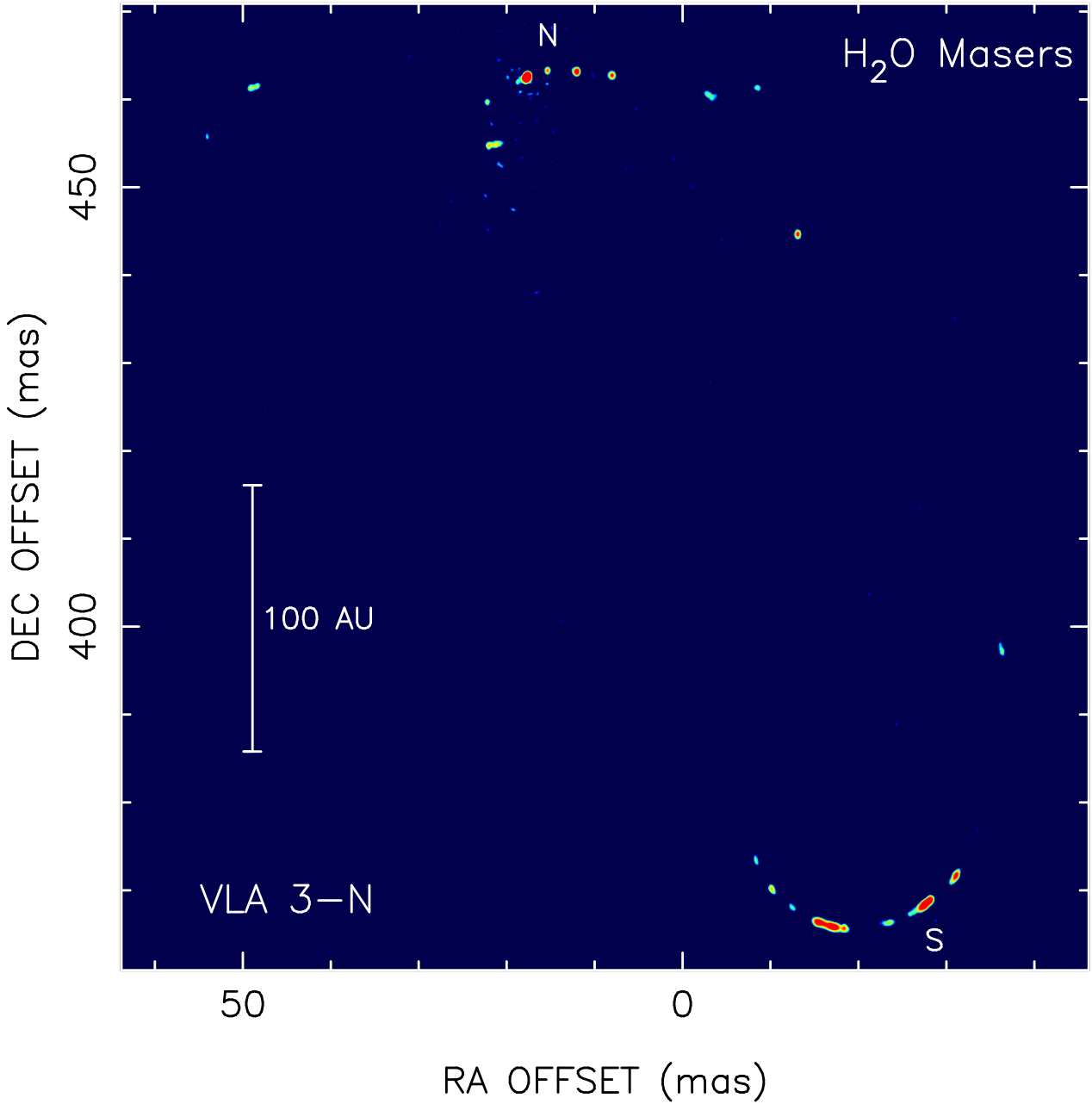} 
 \caption{Integrated intensity image of the water maser emission in the region VLA 3-N (dashed square in Figure 1) for the first epoch of the VLBA observations on 
 2001 Dec 2 (the spatial distribution of the water maser emission is very similar in the two other observed epochs,  2002 Feb 11 and 2002 Mar 5). The colour intensity scale is saturated at 7 Jy~beam$^{-1}$  (beam $\simeq$ 0.45 mas) to show the full water emission structure. The intensity of all the water masers spots is listed in Table A1 (published online).} 

\end{figure*}

\begin{figure*}
 \centering
 \includegraphics[width=168mm]{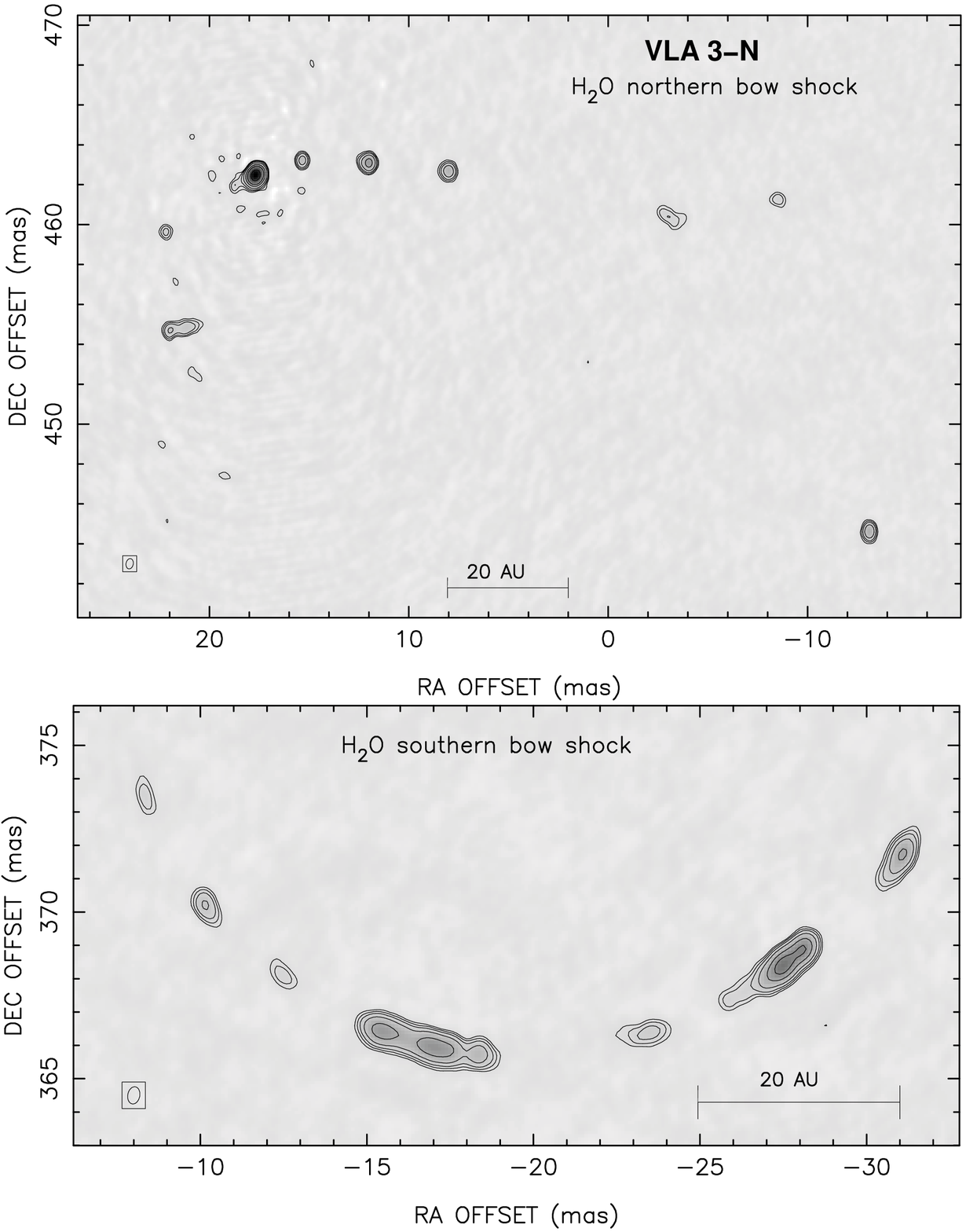} 
 \caption{Close up of the integrated water maser emission
  in VLA 3-N (shown in Figure 2)  for the Northern (upper panel) and Southern (lower panel) 
  bow-shock structures for the first epoch of the VLBA observations on 2001 Dec 2 (the 
  spatial distribution of the water maser emission is very similar in the two other 
  observed epochs,  2002 Feb 11 and 2002 Mar 5). Contours levels are 
 0.7, 1.2, 2, 4, 10, 20, 60 Jy~beam$^{-1}$ (beam $\simeq$ 0.45 mas).}

\end{figure*}

In Table A1 (published online) we list the positions, LSR radial velocity, and intensity
of all the maser spots detected in the region VLA~3-N  for the three observed epochs. 
From this list  of maser spots (Table A1) we identified maser features for proper motion 
measurements purposes. We refer to a maser feature as a group of three or more maser 
spots coinciding within a beam size ($\sim$ 0.4 mas), and each of them appearing in 
consecutive velocity channels (0.21~km s$^{-1}$).  
Table 1 contains the main parameters of the 33 maser features from which we measured proper motions using the three observed epochs (02Dec2001, 11Feb2002, and 05Mar2002). 
For the alignment of our three observed epochs with the data set of Sanna et al (2012) to make a proper comparison (main goal of this work), we identify a maser feature clearly persisting in the two sets of data: S~17 listed by Sanna et al. (2012) (which has a relatively small proper motion), identified in our 
data set as maser feature ID28 (both having similar radial velocities; Tables 1 and 2). We have corrected the positions of the feature ID 28 as a function of time (and therefore to all our data set of 2001-2002), extrapolated from the position and proper motion of the maser S17 (Sanna et al. 2012), assuming that it has moved with constant velocity through the time span of $\sim$ 7 yr. After this alignment, proper motions listed in Table 1 were estimated by a linear fitting to position of the masers of the three different epochs as a function of time. The fact that after this alignment, the whole maser structure of our observations is within that reported by Sanna et al. (2012), and that the estimated shift in position of our observed masers for a time span of seven years coincide with 
the angular separation between the structures observed in 2001-2002 (this paper) and those of 2008-2009 (Sanna et al. 2012) (see below, \S 3) gives our alignment and proper motion estimates an additional measure of robustness.

We note that the maser feature ID~28 is not the maser used to self-calibrate our data, which is associated with the radio continuum source 
VLA~3 ($\sim$ 0.5$''$ south of VLA~3). All the offset positions of the masers given in this paper (tables and figures) are relative to the maser
spot position (0,0) used for self-calibrating the
data of our first epoch of observations (02Dec2001; see above). For comparison purposes, further analysis, and full discussion of all the results
obtained in this region, we also list in Table 2 the 19 water maser features where proper motions have been  measured by Sanna et al. (2012) for the 2008-2009 epochs in VLA~3-N (all the positions listed in Table 2 are also given with respect to our reference position [0,0],  $\alpha$ [J2000] = 20$^h$29$^m$24.879$^s$, $\delta$ [J2000]=
40$^{\circ}$11$'$19.47$''$ $\pm$ 0.01$''$).

\section{Observational results}

In Figure 2 we present an integrated intensity image of the water maser emission obtained 
from our VLBA data toward the region VLA 3-N. The emission shows the presence of two 
shell-like structures of water masers with sizes $\sim$ 30~mas ($\sim$ 100~AU) that resemble bow 
shocks, named hereafter as the {\it Northern} and {\it Southern} bow-shock structures.
These structures are observed in our three epochs of observations (2001-2002). 
The Northern and Southern bow shocks are separated from each other by $\sim$ 100 mas
($\sim$ 330 AU). It is worth noting that the spatial distribution of these water maser 
structures agrees very well with those traced by the VLBA observations reported by Sanna
et al.  (2012), made seven years after our observations (see below). 
This result implies that both the surrounding interstellar medium and the exciting 
source(s) of the masers have been relatively stable over this temporal scale of seven years.

The Northern and Southern bow shocks detected with our VLBA data present significant 
morphological differences. In fact, while the Northern bow shock has a somewhat irregular 
morphology, the Southern one has a remarkable smooth morphology (not so evident in the 
data presented by Sanna et al. 2012), which has remained stable during our three epochs 
of VLBA observations. This is more clearly seen in Figure 3, where we show, as close ups, 
the integrated intensity contour maps of the two bow-shock structures.

The water maser emission in VLA 3-N spans a velocity range from $V_\mathrm{LSR}\simeq -13$ 
to $-2$~km~s$^{-1}$, and flux densities range from $\sim 0.05$ to 30~Jy~beam$^{-1}$ 
(Tables A1 and 1). Most of the water maser emission appears blueshifted with respect 
to the systemic velocity of the extended ambient molecular cloud of the AFGL~2591 massive 
star formation region (V$_{LSR}$ $\simeq$ $-6$~km~s$^{-1}$; van der Tak et al.\ 1999). 
However, the center of the velocity range covered by the masers is closer to the LSR 
velocity of the HC$_3$N line emission from the hot molecular core in VLA 3 
($V_{\rm LSR}\simeq -7$ to $-8$ km s$^{-1}$), measured with the SMA with angular 
resolution $\sim$ $0.4''$ (Jim\'enez-Serra et al. 2012). 
The water maser spectrum of the whole VLA~3-N region is dominated
by two main velocity components ($\sim$ 40~Jy), 
at $V_\mathrm {LSR}\simeq -9.0$ and -7.5 km s$^{-1}$, arising from the Northern and 
Southern bow-shock structures, respectively (see Figure 4).

 \begin{figure}
\centering
 \includegraphics[width=84mm]{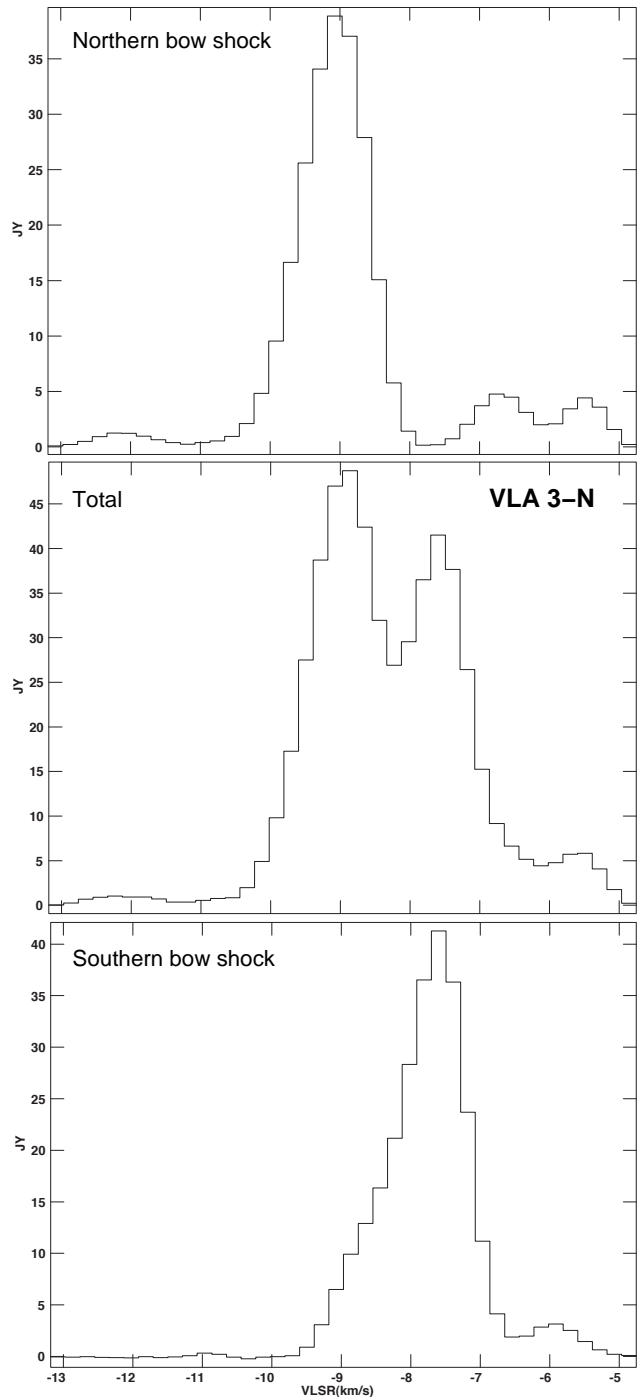} 
 \caption{Water maser spectra of the Northern (top panel) and Southern (bottom panel) 
bow-shock structures observed with the VLBA toward VLA 3-N (see Figures 1 and 2)
on 2001 December 02 (the spectra in the other two observed epochs are similar).
The total spectrum obtained for the full region VLA 3-N ($\sim$ 0.1$''$ size) 
is also shown (central panel).}
\end{figure}

 We measured the proper motions of 33 water maser features located in the VLA~3-N region 
(Table 1), resulting in average tangential velocities of  $\sim$ 20~km~s$^{-1}$ 
($\sim$ 1.3~mas~yr$^{-1}$). In Figure 5 we plotted the positions of all the water 
maser spots measured with the VLBA at epochs 2001-2002 (Table A1), together with 
the proper motion vectors of these 33 maser features (Table 1). We also plotted in 
this figure the 19 maser features measured by Sanna et al. (2012) in 2008-2009, 
as well as their proper motions. The color scale of the symbols represent their LSR radial
velocities.

\begin{figure*}
 \centering
 \includegraphics[width=168mm]{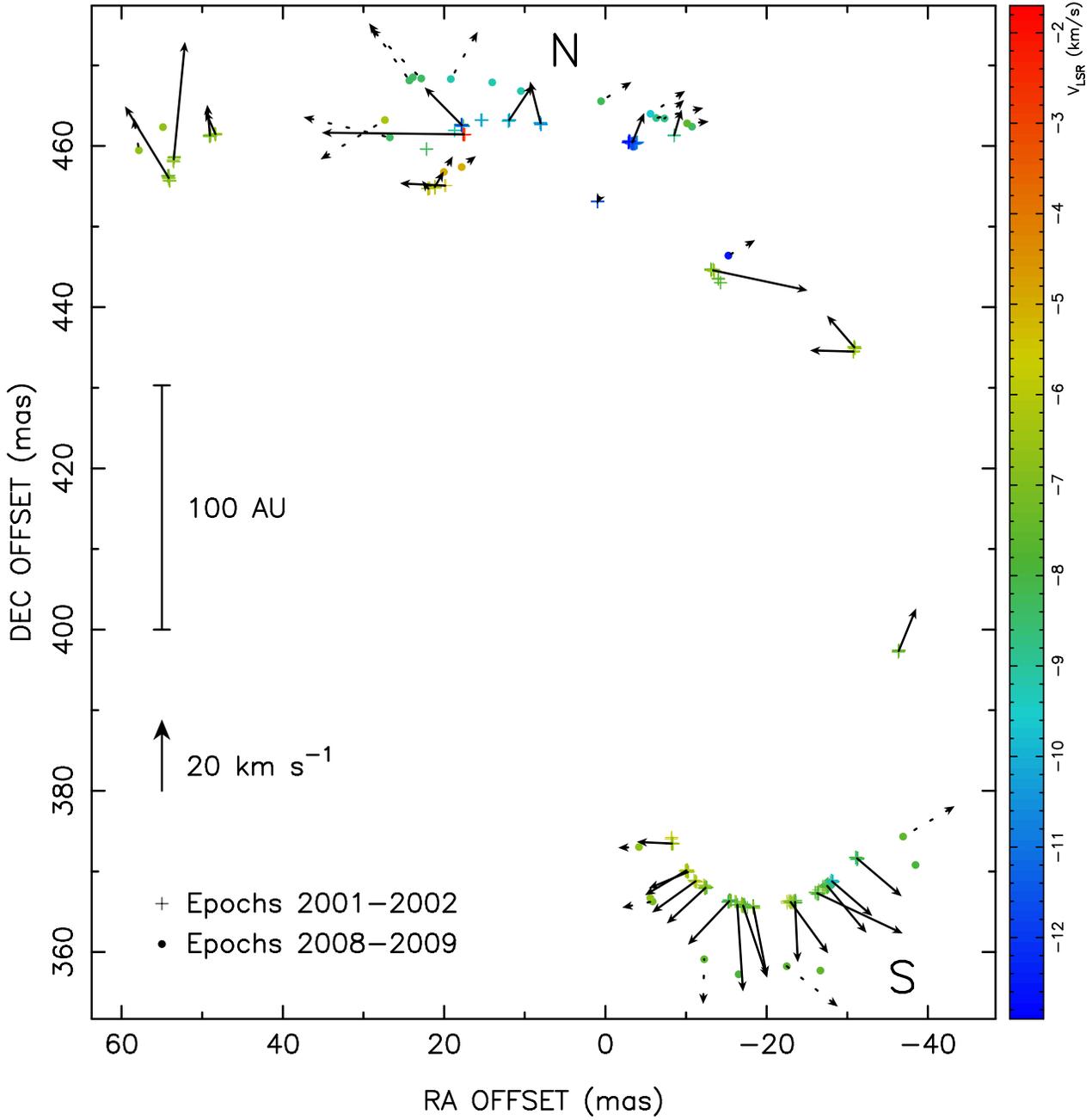}
 
\caption{Positions of all the water maser spots measured with the VLBA in 
the region VLA~3-N in epochs  2001-2002 (plus signs; this paper, Table A1)
and in epochs 2008-2009 (filled circles; Sanna et al. (2012)). The color scale 
represents the LSR radial velocity of the masers. Arrows (solid and dashed
lines for epochs 2001-2002 and 2008-2009, respectively)
represent the proper motion vectors 
of the maser features listed in Tables 1 and 2.
The length of the arrows corresponds to the 
estimated proper motions over the total time span of seven years
(time span between our VLBA observations and those of Sanna et al. 2012). 
Note that the positions of all the maser spots detected  in 2001-2002
fall within the area circumscribed by the masers detected in 2008-2009.
The Northern and Southern bow-shock structures are separated from each other 
by $\sim$ 100 mas ($\sim$ 330 AU) for epochs 
2001-2002, and $\sim$ 120 mas ($\sim$ 400 AU) for epochs 2008-2009 (see \S\S3,4). }
\end{figure*}

\begin{figure*}
 \centering
 \includegraphics[scale=0.78]{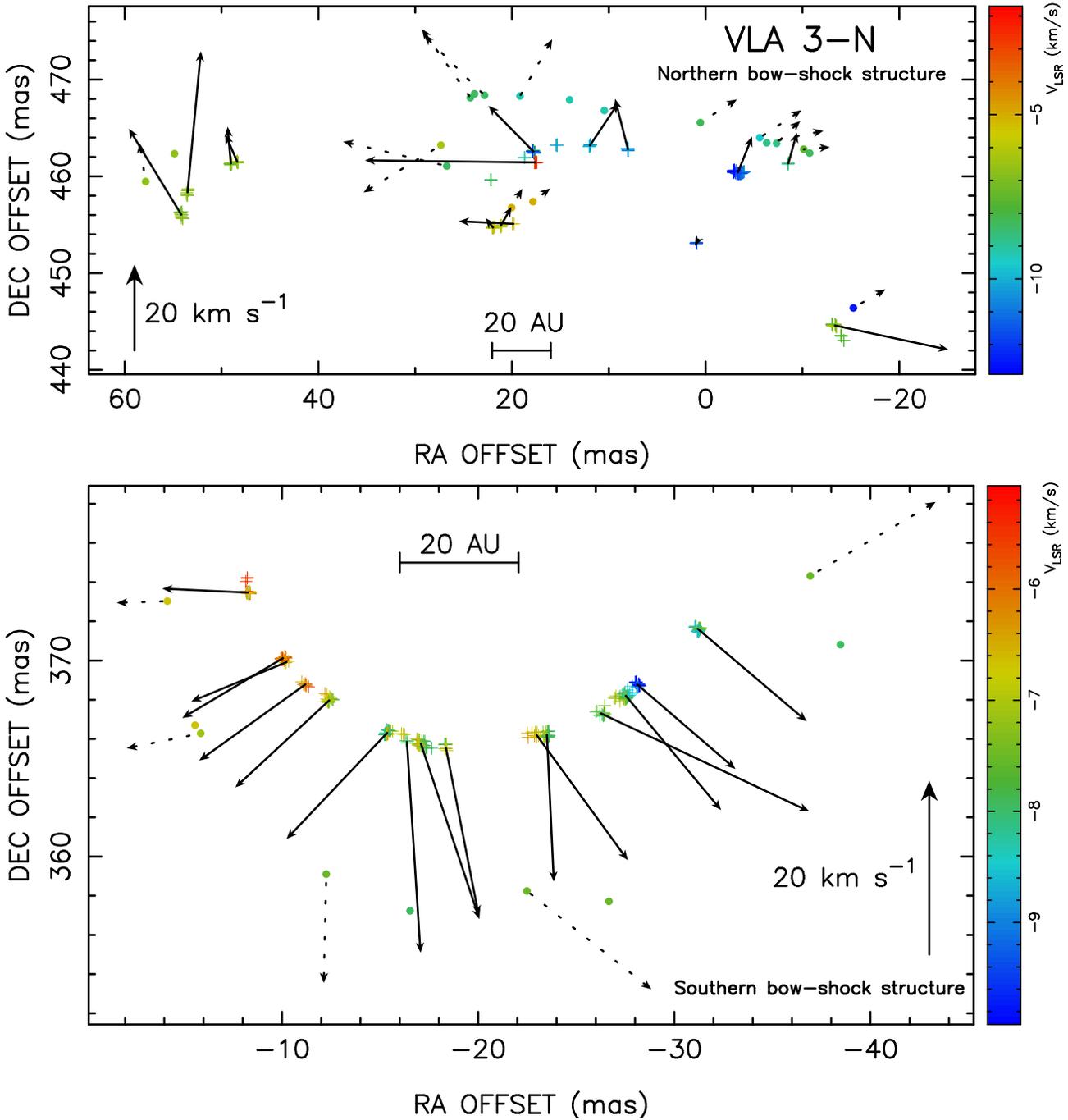} 
 \caption{Same as in Figure 5, but showing a close up of the water maser proper motions of 
the Northern (top panel) and Southern (bottom panel) bow-shock structures in VLA 3-N. 
Note that the lengths of the solid arrows indicating the estimated shift in position for a time span of seven years roughly coincide with the angular separation between the structures observed in 2001-2002 (this paper) and those of 2008-2009 (Sanna et al. 2012), more noticeable for the Southern bow-shock structure. This is consistent with the structures moving at almost constant velocities during the time span of seven years. 
}

\end{figure*}

 \begin{table*}
 \centering
 \begin{minipage}{140mm}
    \caption{Proper motions of the VLBA water maser features in AFGL2591 VLA~3-N (2001-2002 epochs)}
 \begin{tabular}{@{}lrccrrcc@{}}
  \hline
Feature   &  Detected& V$_{LSR}$ &Intensity & $\Delta x$& $\Delta y$ & $V_x$ & $V_y$\\
ID  &  Epochs& (km~s$^{-1}$) & (Jy~beam$^{-1}$) & (mas) & (mas) & (km~s$^{-1}$) & (km~s$^{-1}$)\\
    \hline
    
    1 &   1,2,3 &   $-$6.1  & 0.6  &    $-$8.27    &  373.46 & 9.8 $\pm$ 0.8  &   0.5 $\pm$  0.6\\ 
2 &   1,2,3 &    $-$5.7 &  1.2  &    $-$10.05  & 370.12  & 11.6 $\pm$  0.3&  $-$6.9 $\pm$ 0.2\\
3 &    1,2,3 &    $-$6.5&   0.1 &     $-$10.23 & 369.92  & 10.9 $\pm$ 0.2 &  $-$4.5 $\pm$ 0.3\\
4 &     ~ 2,3 &    $-$6.1&   0.1 &     $-$11.19 &  368.78 & 12.2  &  $-$8.8 \\ 
5 &    1,2,3 &    $-$7.1&   0.4 &     $-$12.41 &  367.97 & 10.8 $\pm$ 1.3 & $-$10.0 $\pm$ 1.1\\ 
6 &    1,2,3 &    $-$7.5&   3.8 &     $-$15.37 &  366.31 & 11.6 $\pm$ 0.2 & $-$12.2 $\pm$ 0.2\\
7 &   2,3   &   $-$7.8 &  1.5 &    $-$16.37  & 365.85  & $-$1.6 &  $-$24.2 \\
8 &    1,2,3 &    $-$7.3&   4.3&     $-$17.08 &  365.76 & $-$6.7 $\pm$ 3.1 & $-$20.2 $\pm$ 1.6\\
9 &    1,2,3 &    $-$7.3&   1.7 &     $-$18.36 &  365.54 & $-$3.8 $\pm$ 1.9 & $-$19.2 $\pm$ 1.1\\
10 &   1,2,3 &    $-$6.7&   0.3&     $-$22.98 &  366.20 & $-$10.5 $\pm$  2.3&  $-$14.4 $\pm$ 0.5\\
11 &   1,2,3 &    $-$7.4&   0.8 &     $-$23.51 &  366.22 & $-$0.8 $\pm$ 0.2 & $-$16.9 $\pm$ 0.2\\
12 &   1,2,3 &    $-$7.7&   0.4 &     $-$26.26 &  367.30 &  $-$23.9 $\pm$ 0.3&  $-$11.3 $\pm$ 0.3\\
13 &   1,2,3 &    $-$8.0 &   6.6 &    $-$27.53 &  368.20 & $-$10.9 $\pm$ 1.1& $-$13.1 $\pm$ 0.9\\
14 &   1,2,3 &    $-$8.9&  5.1 &    $-$28.12 &  368.77 &$-$11.3 $\pm$ 0.1 & $-$9.7 $\pm$ 0.1\\
15 &   1,2,3 &    $-$8.0 &  2.8  &    $-$31.20 &  371.62 & $-$12.5 $\pm$  0.3 & $-$10.6 $\pm$ 0.5\\
16 &   1,2,3 &    $-$7.3&  0.7  &    $-$36.38 &  397.36 & $-$4.8 $\pm$ 0.3  & 11.9 $\pm$ 0.3\\
17 &    2,3 &    $-$6.3 &  0.2  &    $-$30.78 &  434.50 & 12.2  &  0.3  \\
18 &   1,2,3 &    $-$6.5 &  0.2  &    $-$30.88 &  435.01 &  7.7 $\pm$ 0.8 &  8.9  $\pm$ 1.7\\
19 &   1,2,3 &    $-$6.4 &  1.5  &    $-$13.36 &  444.58 & $-$26.4 $\pm$  0.9 &  $-$5.6 $\pm$ 0.2\\
20 &   1,2,3 &   $-$11.5 &  0.1  &      0.96 &  453.10 &  0.3 $\pm$ 0.2  & $-$0.6 $\pm$ 0.3\\
21 &   1,2,3 &   $-$11.9 &  0.3  &     $-$3.35 &  460.43 & $-$3.3 $\pm$ 1.7  &  8.3 $\pm$ 0.6\\
22 &   1,2,3 &    $-$8.7 &  0.5  &     $-$8.54 &  461.33 & $-$2.0 $\pm$ 0.6 &  7.2 $\pm$ 1.3\\
23 &   1,2,3 &    $-$9.8 &  2.0 &      8.03 &  462.80 &  3.0 $\pm$ 0.2 & 11.7 $\pm$ 0.6\\
24 &   1,2,3 &    $-$9.1 &  4.3  &     11.94 &  463.19 & $-$6.6 $\pm$ 1.1  &  9.8 $\pm$ 0.6\\
25 &   1,2,3 &    $-$9.0 & 20.6  &     17.78 &  462.57 & 10.5 $\pm$ 0.1  & 10.6 $\pm$ 0.6\\
26 &    2,3 &    $-$2.3 &  0.4  &     17.56 &  461.44 & 39.7  &  0.5 \\
27 &   1,2 ~ &    $-$5.2 &  0.1  &     19.86 &  455.09 & 12.7   &  0.6 \\
28 &   1,2,3 &    $-$5.3 &  1.2  &     21.16 &  454.90 & $-$2.5  &  4.4 \\
29 &   1,2,3 &    $-$5.4 &  1.3  &     21.96 &  454.71 &  1.7 $\pm$ 0.1  &  2.2 $\pm$ 0.3\\
30 &   1,2,3 &    $-$6.5 &  0.7  &     48.37 &  461.49 &  2.8 $\pm$ 0.2  &  6.4 $\pm$ 0.2\\
31 &   1,2,3 &    $-$6.7 &  0.8  &     49.03 &  461.31 &  0.8 $\pm$ 1.9  &  8.6 $\pm$ 0.5\\
32 &   1,2,3 &    $-$6.8 &  0.1  &     53.52 &  458.35 & $-$3.1 $\pm$ 1.6  & 32.8 $\pm$  3.1\\
33 &   1,2,3 &    $-$6.7 &  0.4  &     54.16 &  456.04 &  12.2 $\pm$ 0.5 &  20.0 $\pm$  0.8\\

    \hline
\end{tabular}

\noindent {\bf Notes}. The water maser features where proper motions have been measured are
numbered in column 1 for identification purposes.  For each feature, the different 
epochs of detection are given in column 2 (Epoch 1: 2001 Dec 02; Epoch 2: 
2002 Feb 11; Epoch 3: 2002 Mar 05). Columns 3 and 4 give, respectively, the LSR 
radial velocity and
peak intensity of the maser features as observed in the first epoch
of detection. The position offset of the features (referred to the
mean epoch of the epochs where the features are detected) are
given in columns 5 and 6, while their proper motions are reported in
columns 7 and 8. All the  offset positions are relative to the maser
spot position (0,0) used for self-calibrating the
data of the first epoch, with absolute position RA(J2000) = 20$^h$29$^m$24.879$^s$, DEC(J2000)=
40$^{\circ}$11$'$19.47$''$ ($\pm$ 0.01$''$) and V$_{LSR}$ = $-$18.8 km~s$^{-1}$. This maser spot used for self calibrating the data
is associated with the radio continuum source VLA 3 (Trinidad et. al. 2003), $\sim$ 0.5$''$ south from the region subject of the study in this paper. For the alignment of the different
observed epochs we used the maser feature S17 observed by Sanna et
al. (2012) (see Table 2) and identified in our data set with maser feature ID~28 (this table), assuming it has constant velocity through the time span of $\sim$ 7 years
between our observations (2001-2002 epochs) and those of Sanna et
al. (2008-2009 epochs). Consequently, for the maser feature ID 28, and for those detected only in two epochs (IDs 4,  7, 17, 26, and 27), there are not uncertainty estimates for the proper motion values (linear fitting). We estimate
that the accuracy in the relative positions of the maser features is $\sim$ 
0.10 mas. The positions, LSR radial velocity, and intensity
of all the maser spots observed in this region in the three epochs, are listed in Table A1 (published on-line; see \S 2) and represented in Figures 5 and 6.
\end{minipage}
\end{table*}

\begin{table*}
 \centering
 \begin{minipage}{140mm}
   \caption{Proper motions of the VLBA water maser features in AFGL2591 VLA~3-N (2008-2009 epochs)}
 \begin{tabular}{@{}lrccrrcc@{}}
  \hline
Feature   &  Detected & V$_{LSR}$ &Intensity & $\Delta x$& $\Delta y$ & $V_x$ & $V_y$\\
ID   &  Epochs& (km~s$^{-1}$) & (Jy~beam$^{-1}$) & (mas) & (mas) & (km~s$^{-1}$) & (km~s$^{-1}$)\\
    \hline
    
S17  &  a,b,c,d   &   $-$5.0  &22.2     &   20.04  & 456.77  & $-$2.5 $\pm$ 1.8  &    4.4 $\pm$  2.1\\    
S18  &  a,b,c,d   &   $-$9.2  & 6.0     &   19.17  & 468.30  & $-$7.5 $\pm$ 1.7  &   13.2 $\pm$ 2.0\\  
S19  &  a,b,c,d   &  $-$12.6  & 5.4     &  $-$15.26  & 446.41  & $-$7.5 $\pm$ 2.0  &    4.3 $\pm$ 2.5\\   
S20  &  a,b,c,d   &   $-$7.5  & 4.5     &  $-$12.26  & 359.10  &  0.3 $\pm$ 2.2  &  $-$12.5 $\pm$ 2.4\\    
S21  &  a,b,c,d   &   $-$8.8  & 2.9     &   $-$6.31  & 463.47  &$-$13.2 $\pm$ 2.1  &    2.8 $\pm$ 2.1\\    
S22  &  a,b,c,d   &   $-$8.8  & 2.3     &   $-$7.33  & 463.42  & $-$5.5 $\pm$ 1.7  &    5.2 $\pm$ 2.0\\    
S24  &  a,b,c ~   &   $-$8.4  & 2.0     &   24.30  & 468.13  & 11.3 $\pm$ 4.2  &   16.0 $\pm$ 4.3\\    
S26  &  a,b,c,d   &   $-$7.5  & 1.6     &  $-$22.49  & 358.24  &$-$14.3 $\pm$ 1.8  &  $-$11.3 $\pm$ 2.2\\    
S28  &  a,b,c,d   &   $-$7.1  & 0.7     &   $-$5.86  & 366.28  &  8.5 $\pm$ 1.8  &   $-$1.7 $\pm$ 2.1\\    
S30  &  a,b,c,d   &   $-$5.0  & 0.5     &   17.83  & 457.38  & $-$4.0 $\pm$ 2.0  &    3.1 $\pm$ 2.0\\    
S32  &  a,b,c,d   &   $-$8.3  & 0.4     &    0.54  & 465.56  & $-$8.5 $\pm$ 1.8  &    5.4 $\pm$ 2.2\\    
S33  &  a,b,c,d   &   $-$9.6  & 0.4     &   $-$5.59  & 464.00  & $-$9.7 $\pm$ 2.1  &    6.4 $\pm$ 2.2\\    
S34  &  a,b,c  ~  &   $-$8.4  & 0.4     &   22.84  & 468.37  & 14.1 $\pm$ 3.3  &   14.0 $\pm$ 8.7\\    
S35  &  a,b,c,d   &   $-$7.5  & 0.4     &  $-$36.93  & 374.32  &$-$14.4 $\pm$ 1.8  &    8.5 $\pm$ 2.2\\    
S36  &  ~ b,c,d    &   $-$8.4  & 0.3     &   26.74  & 461.06  & 24.1 $\pm$ 3.0  &    5.7 $\pm$ 3.2\\    
S37  &  a,b,c ~    &   $-$6.3  & 0.2     &   27.34  & 463.22  & 18.0 $\pm$ 3.3  &  $-$11.0 $\pm$ 4.7\\    
S38  &  a,b,c,d   &   $-$6.7  & 0.2     &   57.85  & 459.46  &  1.4 $\pm$ 1.8  &    9.0 $\pm$ 2.2\\    
S40  &  a,b,c,d   &   $-$6.7  & 0.2     &   $-$4.16  & 373.03  &  5.8 $\pm$ 1.8  &   $-$0.2 $\pm$ 2.1\\    
S41  &  ~ b,c,d    &   $-$7.5  & 0.2     &  $-$10.13  & 462.80  & $-$6.1 $\pm$ 3.7  &    0.5 $\pm$ 3.0\\ 
    \hline
\end{tabular}

\noindent {\bf Notes}. Same as Table 1, but for the water maser features with proper motions measured by Sanna et al. (2012) for the epochs {\it a} (2008 Nov 10), {\it b} 
(2009 May 06), {\it c} (2009 May 13), and {\it d} (2009 Nov 13).
The ID numbers of the features are the same as those listed by Sanna et al. (2012) 
preceded by an S. All the  offset positions are with respect to the same (0,0) position given for the data of Table 1 after alignment of all the observed
epochs (2001-2002 and 2008-2009; see notes in Table 1 and \S 2). In this way the offset positions listed in columns 5 and 6 of this table where obtained
by adding ($-$631.01 mas, +151.08 mas) to the ($\Delta x$, $\Delta y$) positions listed in columns 5 and 6 of Table 2 of Sanna et al. (2012).
\end{minipage}
\end{table*}

Figure 5 shows that the northern masers of VLA~3-N  are moving essentially northward,
while the southern masers are moving southward (epochs 2001-2002). This is fully consistent with the results reported by  Sanna et al. (2012) for epochs 2008-2009. In addition, we found that 
the two main water maser structures detected in epochs 2001-2002 in VLA~3-N (i.e.,
the Northern and Southern bow shocks) fall within the corresponding structures observed 
in epochs 2008-2009. In fact, the Northern and Southern bow-shock structures are separated 
from each other by $\sim$ 100 mas ($\sim$ 330 AU) in epochs 2001-2002, while they are separated $\sim$ 120 mas ($\sim$ 400 AU) in epochs 2008-2009. This implies that the two bow-shock 
structures have increased their angular separation in the time span of seven years at a 
relative proper motion of $\sim$ 2.8~mas~yr$^{-1}$. If we assume expansion from a common center, this would correspond to expanding proper motions of $\sim$ 1.4~mas~yr$^{-1}$ 
(equivalent to a velocity of $\sim$ 22~km~s$^{-1}$). This velocity is consistent with the proper motion values we have measured for the individual water maser features. 
This is more clearly seen from Figure 6, where we show as a close up the proper motions 
in the Northern and Southern bow-shock structures. The magnitudes
 of individual proper 
motions, represented by the arrow lengths, roughly coincide with the rate of increase in
angular separation between the structures observed in 2001-2002 and those of 2008-2009. 
This is especially remarkable for the Southern bow-shock structure.

All these characteristics are consistent with the structures moving at almost constant 
velocities during a time span of seven years. Considering that the difference between 
the radial velocities of the Northern and Southern bow-shock structures is a few 
km~s$^{-1}$ (Figure 4), and their radial velocities are relatively close to the radial 
velocity of the HC$_3$N line emission peak observed $\sim$ 0.25$''$ north from VLA~3 
within the hot molecular core (Jim\'enez-Serra et al. 2012), together with the large 
tangential velocities that we obtain ($\sim$ 20~km~s$^{-1}$), we conclude that the water
masers structures in VLA~3-N are likely moving almost on the plane of the sky. 
In addition, within each of the two bow-shock structures, there are also small radial velocity differences. For the Southern bow shock, a small velocity shift of $\sim$ 2~km~s$^{-1}$ 
is observed, increasing west to east from V$_{LSR}$ $\simeq$ $-8$ km~s$^{-1}$ to 
V$_{LSR}$ $\simeq$ $-6$ km~s$^{-1}$ (Figure 6). For the Northern bow shock, radial velocity differences are relatively larger than in the Southern one, up to $\sim$ 4-5~km~s$^{-1}$, but these differences are distributed more irregularly.

\section{Discussion: The maser shell as a stellar wind bow shock}

The Northern and Southern shell-like structures traced by the water maser emission
in VLA~3-N show systematic expanding motions that could imply one or two different
centers of star formation activity.
Here we consider two possible scenarios to explain the spatio-kinematical distribution 
of these masers: i) a static central source driving the two bow-shock structures; 
ii) two independent driving sources, one of them exciting the Northern 
bow-shock structure, and the other one, a young runaway star moving in
the local molecular medium at $\sim$ 20~km~s$^{-1}$, exciting and molding 
the smoother Southern bow-shock structure. Given that the Southern structure
has a remarkably smooth morphology and systematic proper motions, we 
use this structure to develop the model.

\subsection{The bow-shock model}

The water masers of the Southern structure
detected in the VLA 3-N region trace a more or less
smooth arc on the plane of the sky. When compared with the structure
observed by Sanna et al. (2012), it is clear that the arc-like structure
(traced by the masers) is moving as a whole, in an approximately southward
direction.

Sanna et al. (2012) interpreted this arc of masers, together with the
Northern structure, as part of an elliptical
bow shock expanding into the molecular
cloud, driven by a static central source, and expanding into the 
molecular cloud. 
The heating and compression of the molecular material at the shock
gives rise to the observed maser emission. 

We think that this bow-shock interpretation is basically correct. But here we present a generalized and more detailed model for this bow-shock scenario.
Our interpretation is based on the work of Raga et al. (1997) on the
proper motions of condensations in a bow-shock flow. This work presents
a derivation of  the basic parameters of the bow shock
(such as its velocity, direction and environmental motion) from
a set of observed proper motions of water maser features.

Firstly, we assume that the bow shock has an axially symmetric shape and
that it moves with a constant velocity (in magnitude and direction)
with respect to the surrounding molecular cloud. As the bow shock moves,
it heats, compresses and pushes the molecular cloud material, producing
a thin shell of emitting gas.

The bow shock moves at a velocity $v_{bs}$, directed at an angle
$\phi$ with respect to the plane of the sky. The water masers are
likely to trace the bow shock ``edge'' (i.e., the region in which
the thin shell is tangential to the line of sight), since this is the line
of maximum (local) optical depth as seen by the observer.

If we assume that the bow shock
has a paraboloidal shape, the projection of the bow shock
(which moves in an arbitrary direction) on the plane of the sky
is also a parabola. This is demonstrated in Appendix A, in which we
develop the equations describing the shape of the bow shock.

\begin{table}
 \caption{Parameters of the parabolic fits to the bow shapes}
 \label{tabfits}
 \begin{tabular}{@{}lccccc}
  \hline
Epoch & $t$  & $x_0$  & $y_0$ & $\theta$ & $B$ \\
      &   (days)   &   (mas)     &    (mas)    &    ($^\circ$)      & (mas$^{-1}$) \\         
  \hline
1 & 0 & $-$17.68 & 365.68 & $-$9.73 & 0.0536 \\
2 & 71 & $-$17.64 & 365.44 & $-$9.90 & 0.0532 \\
3 & 93 & $-$17.72 & 365.32 & $-$9.45 & 0.0532 \\
S$^1$  & 2535 & $-$19.50 & 356.56 & $-$2.18 & 0.0530 \\
$\sigma$$^2$ & $\ldots$ & 1.80 & 0.39 & 5.27 & 0.0038 \\
  \hline
 \end{tabular}
 
$^1$ Sanna et al. (2012). \\
$^2$ Standard deviations of the values obtained in the four parabolic fits.
\end{table}

\begin{table}
 \caption{Parameters of the bow shock for different orientations}
 \label{tabmod}
 \begin{tabular}{@{}cccc}
  \hline
$\phi$  & $R_0$  & $v_{bs}$  & $R'_0$ \\
  ($^\circ$)      &   (AU)    &   (km s$^{-1}$)   &   (mas)  \\  
  \hline
0 &  18.6 & 24.0 & 5.6 \\
30 & 16.1 & 27.7 & 5.4 \\
45 & 13.1 & 33.9 & 5.2 \\
60 & \phantom{0}9.3  & 48.0 & 4.9 \\
90 & \phantom{0}0.0  & $\infty$ & 4.7 \\
  \hline
 \end{tabular}
\end{table}

\subsection{Fits to the observed bow shapes}

We now use equations (\ref{xyp}) to (\ref{ypxy}) to carry out
an unweighted least squares fit to the $(x,y)$ positions of the observed
masers of the Southern structure
(see Table A1), from which we obtain the
parameters $(x_0,y_0,\theta, B)$ of the best-fit parabola.
Fits are done to our three observed epochs, as well as to
the 2008-2009 maser positions of Sanna et al. (2012).

\begin{figure}
\centering
\includegraphics[width=84mm]{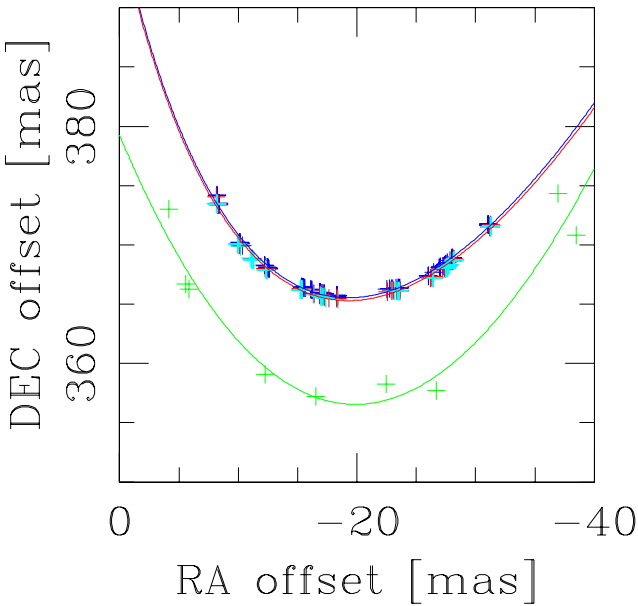}
\caption{Maser positions from our three epochs (blue, magenta,
and red crosses, see Table A1) and the maser positions from Sanna et al. (2012)
(green crosses, see Table 2). The parabolic fits to the four sets
of maser positions are shown with solid lines of the appropriate colours.}
\label{a2}
\end{figure}

The fits are shown in Figure \ref{a2}, and the
resulting parameters (together with their
errors) are given in Table \ref{tabfits}. 
As seen from the standard deviation of the values obtained for each parameter
in the fits of the different epochs, quoted in the bottom line of this table,
the only parameters that show significant trends with time are
$x_0$ and $y_0$ (i.e., the position of the head of the projected bow shock),
while the dispersion of values of theta and B is small, consistent with a 
time-independent behavior.

\begin{figure}
\centering
\includegraphics[width=84mm]{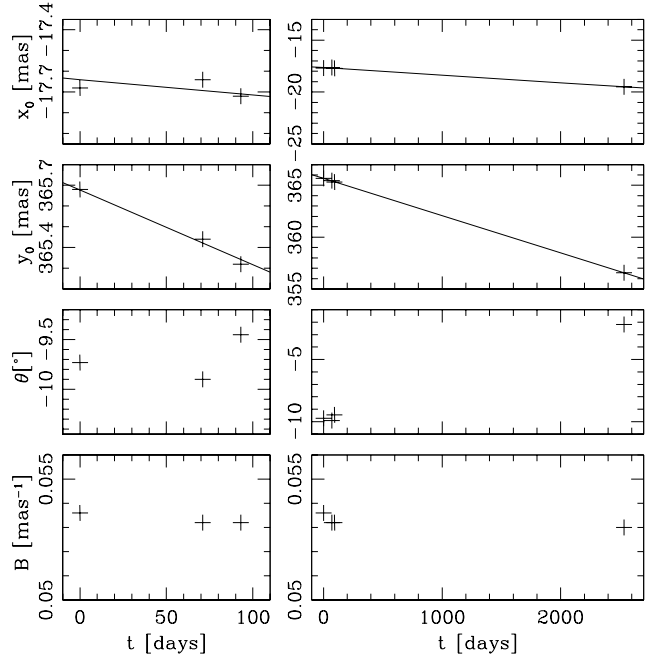}
\caption{Parameters from the parabolic fits to the masers
of our three epochs and the masers of Sanna et al. (2012)
as a function of time (with $t=0$ corresponding to our first epoch: 2001
Dec 02). We have obtained linear fits for the position $(x_0,y_0)$ of
the tip of the bow shock in the four available epochs,
which are shown with solid lines in the
corresponding plots (the same fits are shown in the plots
of the left and right
columns). The graphs of the right column show the four epochs,
and the left column graphs show our three epochs with an expanded
time-axis.}
\label{a3}
\end{figure}

In Figure \ref{a3}
we plot the values of $x_0$ and $y_0$, $\theta$ and $B$ as a function of time.
For $x_0$ and $y_0$ we have carried out
least squares fits to the resulting time-dependencies (also shown in Figure
\ref{a3}), from which we obtain velocities of
$(-0.27\pm 0.31,\,-1.31\pm 0.07)$~mas yr$^{-1}$
(in $\alpha$ and $\delta$, respectively),
corresponding to $(-4.2\pm 4.8,\,-20.6\pm 1.0)$~km s$^{-1}$
at a distance of 3.3 kpc.
The full plane of the sky velocity therefore is of $21.0$~km s$^{-1}$,
which is consistent with the proper motion velocities of the fastest
masers (see Tables 1 and 2).

We have to note that a similar fit for the observed masers of the Northern structure
is not possible because this structure shows a somewhat irregular morphology.
However, we think that the nature of this structure is similar to that of the
Southern bow shock and that the observed morphological differences could be
due to a non uniform VLA 3-N ambient medium, being the density or turbulence
larger in the northern parts. We also can not rule out that the Northern
structure is more evolved and it has begun to disperse.

\subsection{The motions of the individual masers}

\begin{figure}
\centering
\includegraphics[width=84mm]{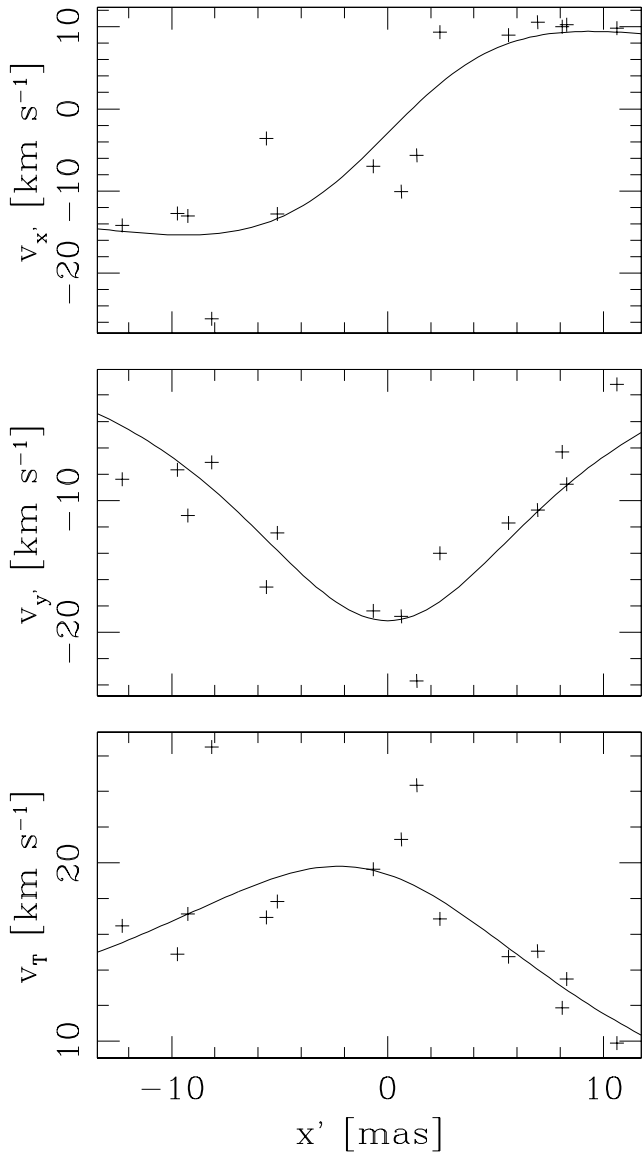}
\caption{ The crosses show the proper motions $v_{x'}$, $v_{y'}$ and
$v_T=\sqrt{v_{x'}^2+v_{y'}^2}$ of the individual maser
spots as a function of the $x'$ coordinate (measured perpendicular to
the axis of the projected bow shapes, see Figure \ref{a1}). The
solid lines correspond to the least squares fits described
in the text.}
\label{a4}
\end{figure}

The motions of masers lying on the projected edge of a bow shock
are also derived in Appendix  A.
We carry out least squares fits of equations (\ref{vxpp}-\ref{vypp})
with $B=0.0533$\,mas$^{-1}$ and $\theta=9.7^\circ$ (the mean values
of the fits to the
bow shapes, see the right column of Table \ref{tabfits})
to the observed $v_{x'}$ and $v_{y'}$ vs. $x'$ dependencies, from which
we obtain the parameters of the bow shock. The corresponding fits
are shown in Figure \ref{a4}. The parameters resulting from the
$v_{x'}$ vs. $x'$ fit are:
\begin{equation}
v^0_{x'}=-(2.95\pm 1.32)\,{\rm km\,s^{-1}}\,\,\,\,\,v_{bs}\cos\phi=(24.80
\pm 3.17)\,
{\rm km\,s^{-1}}\,,
\label{fitx}
\end{equation}
and the parameters resulting from the $v_{y'}$ vs. $x'$ fit are:
\begin{equation}
v^0_{y'}=(4.23\pm 2.55)\,{\rm km\,s^{-1}}\,\,\,\,\,v_{bs}\cos\phi=(23.35
\pm 3.55)\, {\rm km\,s^{-1}}\,.
\label{fity}
\end{equation}
The values of $v_{bs}\cos\phi$ obtained from both fits are consistent.
Also, using the inverted form of equations (\ref{vxp2}) and (\ref{vyp2}),
from the fitted values of $v^0_{x'}$ and $v^0_{y'}$ (see equations
\ref{fitx}-\ref{fity}), we obtain the motion on the plane of the
sky
\begin{equation}
v_x^0=(-3.62\pm 2.72)\,{\rm km\,s^{-1}}\,\,\,\,\,v_y^0=
(3.67\pm 2.75)\,{\rm km\,s^{-1}}\,,
\label{emot}
\end{equation}
of the environment into which the bow shock is traveling. The
$(v_x^0,v_y^0)$ velocity corresponds to the motion of the
molecular cloud with respect to the observer.

If we subtract this environmental motion from the velocity of the
tip of the bow shock ($-4.2$ and $-20.6$ km s$^{-1}$ along the $x$ and
$y$ coordinates, calculated above from the fits to the observed bow
shapes), we obtain a $(-0.57,-24.24)$ km s$^{-1}$ velocity (in
$\alpha$ and $\delta$, respectively) of the bow shock with respect
to its surrounding environment. This corresponds to a full plane of
the sky motion
of 24.25 km s$^{-1}$ of the bow shock with respect to the molecular cloud,
which is consistent with the values of $v_{bs}\cos\phi$ obtained
from the fits to the $v^0_{x'}$ and $v^0_{y'}$ vs. $x'$ dependencies
(see equations \ref{fitx} and \ref{fity}).

In this way, we find that the proper motions derived from parabolic
fits to the bow shapes (in the 4 different epochs) and the proper motions
of the individual masers are all consistent with a single, parabolic
bow shock model with a ``shape parameter'' $B=0.0533$ mas$^{-1}$
(see equation \ref{xyp}) with a plane of the sky motion
of $(-0.6, 24.2)$ km s$^{-1}$ (along $\alpha$ and $\delta$) with
respect to an environment with a
$(-3.6,3.7)$ km s$^{-1}$ plane of the sky motion
(see equation \ref{emot}).

Now, if the maser spots are located in the leading edge of the
projected bow shock, we expect their line of sight velocities to be zero.
The observed small LSR velocity shift of the masers in the Southern bow-shock structure, with velocities between $-8$ and $-6$ km s$^{-1}$ (see \S 3.1 and Table 1), therefore indicates that
the environment molecular gas has radial velocities in this range. This result
is in agreement with the millimeter CO and submillimeter HC$_3$N observations
of van der Tak et al. (1999) and Jim\'enez-Serra et al. (2012), respectively.

\subsection{Parameters of the stellar wind: the runaway star scenario}

The fact that the Northern and Southern bow-shock structures detected with our data present significant morphological differences, with the Southern one showing a remarkably smoother morphology, led us to consider an alternative scenario in which we have two embedded driving sources in VLA 3-N (instead of a single static central driving source). One exciting the most irregular Northern bow-shock structure, and the other one exciting the smoother Southern bow shock.

Let us now assume that the Southern bow shock is produced by the interaction
of a spherical wind from a runaway star with a uniform molecular
environment (while the Northern bow shock is produced by another
different driving source). This problem has been studied,
e.g., by Cant\'o et al. (1996),
who give an analytic expression for the shape of the bow shock. This
shape will remain unchanged as the star moves through the cloud, moving
together with the star (consistent with the fact that the observed
``shape parameter'' $B$ is approximately constant, see Table \ref{tabfits}).

As mentioned above, the shape of the bow shock (close to its apex)
can be approximated by the parabola given by equation (\ref{rwstar}),
with a stagnation radius $R_0$ given by the ram pressure balance
condition
\begin{equation}
\frac{{\dot M}_w v_w}{4\pi R_0^2}=\rho_0 v_*^2\,,
\label{ramp}
\end{equation}
where ${\dot M}_w$ and $v_w$ are the wind mass loss rate and velocity
(respectively), $v_*$ is the velocity of the star with respect to the
cloud and $\rho_0$ is the environmental density. From equations
(\ref{rw}), (\ref{rwstar}) and (\ref{xyp}) it follows that
\begin{equation}
R_0=\frac{3\cos\phi}{10B}\,.
\label{r00}
\end{equation}
Taking the $B=0.0533$ mas$^{-1}$ value deduced from the observed
bow shapes (see Table \ref{tabfits}), we then obtain
\begin{equation}
\left(\frac{R_0}{\rm AU}\right)=18.6\,\cos\phi\,,
\label{r01}
\end{equation}
for a distance of 3.3 kpc.

Now, from equations (\ref{fitx}-\ref{fity}) we have
\begin{equation}
v_{bs}\approx \frac{24.0\,{\rm km\,s^{-1}}}{\cos \phi}\,.
\label{vvv}
\end{equation}
Also, the projected distance between the observed tip of the bow shock and
the star is
\begin{equation}
R'_0=|w'_0|+R_0\cos\phi\,,
\label{rrrp}
\end{equation}
where $w'_0$ is given by equation (\ref{wpp}). Using Eqs.
(\ref{xyp}), (\ref{r00}) and (\ref{rrrp}), we obtain
\begin{equation}
R'_0=\frac{1}{4B}\left(1+\frac{1}{5}\cos^2\phi\right)\,,
\label{rp5}
\end{equation}
or
\begin{equation}
\left(\frac{R'_0}{\rm mas}\right)=4.69\left(1+\frac{1}{5}\cos^2\phi\right)
\label{rp6}
\end{equation}
for $B=0.0533$ mas$^{-1}$. In Table \ref{tabmod}, we give the values
of $R_0$, $v_{bs}$ and $R'_0$ for different values of the orientation
angle $\phi$ between the symmetry axis of the flow and the plane of the
sky, obtained from equations (\ref{r01}), (\ref{vvv}) and (\ref{rp6}).

As expected, for $\phi=90^\circ$ there is no information about
the bow-shock velocity. Interestingly, the position of the
star with respect to the observed
bow-shock tip ($R'_0$) is limited to a range between
4.7 and 5.6 mas (for all values of $\phi$). The narrowness of
this interval is a consequence of the weak dependence of
$R'_0$ on the orientation angle $\phi$ (see equation \ref{rp5}).

To estimate the stellar wind parameters, we have to fix a value for $R_0$
and use equation (\ref{ramp}). Let us take $R_0\approx 15$ AU and $v_{bs}=30$
km s$^{-1}$ (corresponding to $\phi\sim 37^\circ$),
which are representative values for a bow shock with an angle
$\phi$ (with respect to the plane of the sky) not exceeding
$\sim 60^\circ$ (see Table \ref{tabmod}).
This velocity is highly supersonic with respect to the $\leq 1$ km s$^{-1}$
sound speed of the molecular cloud, and the shock is therefore strong.

Since in our model the shock velocity at the bow-shock tip is
$v_{bs}=v_*$, the right hand side of equation (\ref{ramp}) therefore
represents the pressure in the cooling region of the shock. We then
have
\begin{equation}
\rho_0 v_*^2=\rho_0 v_{bs}^2=nkT\,,
\label{sshock}
\end{equation}
where $n$ and $T$ are the density and temperature of the cooling
region (in which the maser emission is produced).

According to Claussen et al. (1997), the density $n(H_2)$ of
molecular hydrogen and the temperature $T$ for the excitation of water maser
emission at 22 GHz must be in the ranges $n(H_2)\sim 10^8$-$10^9$ cm$^{-3}$
and $T\sim 200$-300 K. Therefore, $nkT\sim (0.28$-$4.1)\times 10^{-5}$ dyn
cm$^{-2}$. We can then combine eqs. (\ref{ramp}) and (\ref{sshock})
to obtain:
\begin{equation}
{\dot M}_wv_w=4\pi R_0^2 (nkT)\,,
\label{nkt}
\end{equation}
which for the parameters of our problem becomes
$$\left(\frac{{\dot M}_w}{\rm 10^{-8}M_\odot\,yr^{-1}}\right)
\left(\frac{v_w}{\rm 100\,km\,s^{-1}}\right)=$$
\begin{equation}
0.45 \left(\frac{R_0}{\rm 10\,AU}\right)^2
\left(\frac{nkT}{\rm 10^{-5}dyn\,cm^{-2}}\right)\approx 0.28 - 4.1\,,
\label{nkt2}
\end{equation}
where the second equality corresponds to $R_0=15$ AU and $nkT
\sim (0.28$-$4.1)\times 10^{-5}$ dyn cm$^{-2}$ (as estimated above).
This result indicates that the observed maser arc could be produced
by a young, low mass star with a wind velocity $v_w \sim 100$ km s$^{-1}$
and a moderate mass loss rate ${\dot M_w}\sim {\rm 10^{-8}M_\odot\,yr^{-1}}$.

Finally, let us estimate the density of the surrounding cloud. For this, we
consider an average mass ${\overline m}\approx 2m_H$ for the gas particles.
From Eq. (\ref{sshock}) we then obtain:
$$\left(\frac{n_0}{\rm cm^{-3}}\right)={3.0\times 10^6}\,
\left(\frac{nkT}{10^{-5}}\right)
\left(\frac{\rm 10\,km\,s^{-1}}{v_{bs}}\right)^2$$
\begin{equation}
\approx (0.93 - 13.6)\times 10^5\,,
\label{acaya}
\end{equation}
where for the second equality we have set $v_{bs}=30$ km s$^{-1}$ and
$nkT$ in the range estimated above.

\subsection{Summary}

We find that the proper motions derived from parabolic fits to
the observed Southern arc and the proper motions of the individual masers
are consistent with a bow shock (with a time-independent shape)
moving at a velocity of
$\sim 24$~km s$^{-1}$ within an environment with a velocity
of $\sim 5$~km s$^{-1}$ (both on the plane of the sky).

If we model this bow shock as the interaction of an isotropic
stellar wind with a homogeneous, streaming environment, we find
that the observations are consistent with a wind with a
velocity $v_w \sim 100$~km s$^{-1}$
and a mass loss rate ${\dot M_w}\sim {\rm 10^{-8}~M_\odot\,yr^{-1}}$.
Therefore, the source producing the observed maser arc structure
could be a normal low mass, young star, moving at a velocity of
$\sim 24$~km s$^{-1}$ on the plane of the sky.

Interestingly, the stellar wind bow-shock model predicts that the
position of the star should lie $\approx 4.7 - 5.6$~mas approximately
north of the apex of the bow shock (the range of possible distances
corresponding to different possible
values of the unknown angle $\phi$ between
the bow-shock axis and the plane of the sky).
In an attempt to detect a radio continuum source at the predicted position by the model, we analyzed the VLA archive data of project AJ337.
This project includes observations made in the A (2007 July 26),
B (2008 January 18) and C (2008 March 09) configurations at
3.6~cm and 7~mm. The data were reduced following the NRAO recommendations
and the different configurations were concatenated to obtain
images of good quality. We clearly detect VLA 3 at both wavelengths,
but we fail to detect a source towards VLA~3-N with
3-$\sigma$ upper limits of 0.036~mJy~beam$^{-1}$ ($\lambda$ = 3.6~cm; beam $\simeq$ 0.24$''$) and
0.20~mJy~beam$^{-1}$ ($\lambda$ = 7~mm; beam $\simeq$ 0.15$''$).
Future searches with higher sensitivity
of the emission from this source will show whether or not our
bow-shock model works well in its present form.

It is important to note that a detailed model was not made for the water masers 
tracing the Northern bow-shock structure. As it has been mentioned, these masers are 
not showing a remarkably smooth morphology; instead, they are tracing an structure
with a somewhat irregular morphology, opening the possibility that it is excited by another different YSO and in a possibly more turbulent region.

\section{Conclusions}

We have performed and analyzed multi-epoch VLBA water maser observations toward the high-mass star forming region AFGL 2591 with an angular
resolution of 0.45 mas. Clusters of masers were detected toward the massive YSOs VLA2, VLA 3, and $\sim$ 0.5$''$ ($\sim$ 1700 AU) north of VLA 3, identifying a new center(s) of star formation by its outflow activity (VLA 3-N). The water maser emission in VLA 3-N is distributed in the form of two bow shock-like structures separated from each other by $\sim$ 100~mas ($\sim$ 330 AU). The Northern bow shock-like structure is moving essentially northward, while the Southern one is moving southward. These water maser structures observed in epochs 2001-2002 are contained within 
the area circumscribed by the corresponding structures 
observed by Sanna et al. (2012) in epochs 2008-2009, implying that these structures have persisted during the time span of seven years but increasing their angular separation at a relative proper motion of $\sim$ 2.8~mas~yr$^{-1}$. We find that, while the Northern bow shock-like structure has a somewhat irregular morphology, the Southern one
has a remarkably smooth morphology. We have considered two alternative scenarios to explain these structures: (1) a single, static YSO ejecting a jet, and driving the two bow shock-like structures and (2) at least two YSOs, each of them exciting one of the structures.

In particular, we made a detailed modeling for the Southern structure and found that its spatio-kinematical distribution fits very well with a bow shock with a time-independent shape. We have modeled this bow shock as the interaction of an isotropic stellar wind from a runaway YSO with an homogeneous, streaming environment. The YSO moves at a velocity
 of $\sim$ 24~km~s$^{-1}$ on the plane of the sky.  Our model predicts that the position of this driving YSO (with $v_w \sim 100$ km s$^{-1}$
and ${\dot M_w}\sim {\rm 10^{-8}~M_\odot\,yr^{-1}}$)  should lie $\sim$ 5~mas  north of the apex of the Southern bow shock. 

Future, sensitive radio continuum observations are necessary to discriminate between the different scenarios, in particular to identify the still unseen driving source(s) and 
determine its (their) nature.

\section*{Acknowledgments}

We would like to thank our referee for the very careful and useful report on our manuscript. GA, CC-G, RE, JFG, and JMT acknowledge support from MICINN (Spain) AYA2011-30228-C03 grant (co-funded with FEDER funds). JC and ACR acknowledge support from CONACyT grant 61547. 
SC acknowledges support from CONACyT grants 60581 and 168251.
LFR acknowledges the support of DGAPA, UNAM, and of CONACyT (M\'exico).
MAT acknowledges support from CONACyT grant 82543. RE and JMT acknowledge support from AGAUR (Catalonia) 2009SGR1172 grant. The ICC (UB) is a CSIC-Associated Unit through the ICE (CSIC).


\begin{appendix}

\section{Bow shock model}

\subsection{The shape of the bow shock}

Let $r(w)$ be the cylindrical radius of the bow shock as a function
of the distance $w$ measured along the symmetry axis. Then, the locus
of the edge of the bow shock on the $(x',w')$ plane of the sky
(with $x'=0$, $w'=0$ corresponding to the observed bow-shock tip,
the $x'$ axis perpendicular to the projected symmetry axis, and $w'$
perpendicular to $x'$) is given by (see Raga et al. 1997):
\begin{equation}
x'=r\,\left[1-\left(\frac{dr}{dw}\right)^2\tan^2\phi\right]^{1/2}\,,
\label{yr}
\end{equation}
\begin{equation}
w'=w\,\cos\phi\,\left[1-\frac{r}{w}\left(\frac{dr}{dw}\right)
\tan^2\phi\right]\,.
\label{wp}
\end{equation}

To proceed, let us assume that $r(w)$ is a parabola:
\begin{equation}
r=Aw^{1/2}\,,
\label{rw}
\end{equation}
where $A$ is a constant. We assume this shape for the bow shock because
any ``flat topped'' functional form can be approximated to second
order as a parabola in the region of its apex. For instance, a bow
shock generated by a star with an isotropic wind moving into a uniform
medium has a complex shape (Wilkin 1996; Cant\'o et al. 1996), which
in the apex region can be approximated by
\begin{equation}
r=\left(\frac{10R_0}{3}\right)^{1/2}w^{1/2}\,,
\label{rwstar}
\end{equation}
where $R_0$ is the distance between the star and the tip of the bow shock.

Combining equations (\ref{yr}-\ref{rw}), one can show that the locus
of the edge of the bow shock on the plane
of the sky is also a parabola, given by
\begin{equation}
x'=\frac{A}{\cos^{1/2}\phi}y'^{1/2}\,,
\label{yp}
\end{equation}
where
\begin{equation}
y'=w'-w'_0\,,
\label{zp}
\end{equation}
\begin{equation}
w'_0=-\frac{A^2}{4}\tan\phi \sin \phi\,.
\label{wpp}
\end{equation}
Actually, $w'_0$ is the projected distance between the real tip
of the bow shock and its ``projected head'' on the plane of the sky.

\begin{figure}
\centering
\includegraphics[width=84mm]{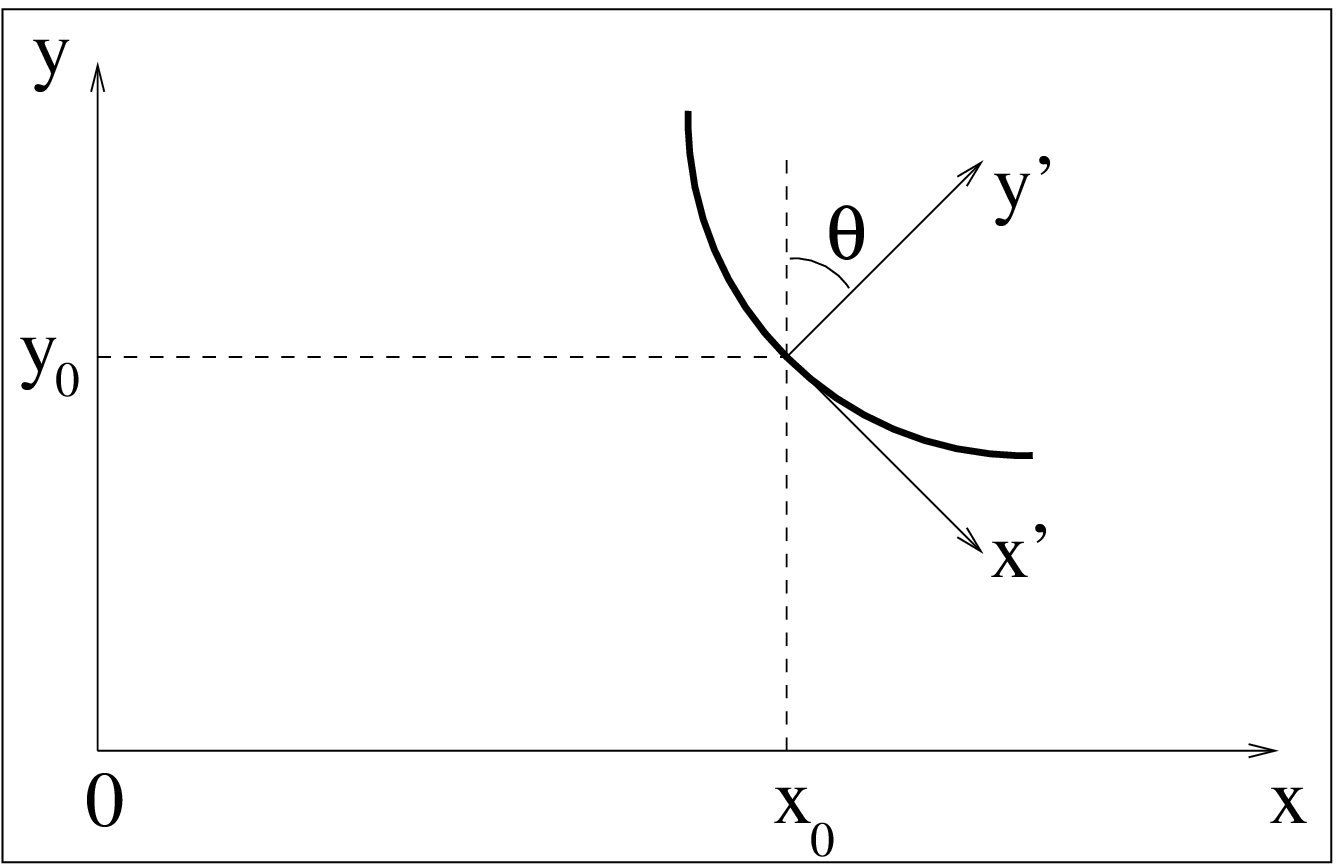}
\caption{Schematic diagram showing the observer's
plane-of-the-sky reference frame $(x,y)$ and the
$(x',y')$ reference frame aligned with the projected
bow shock shape (with the $y'$ axis parallel to the symmetry axis
of the bow shock, and $y'=0$ at the projected tip of the bow shock).
The position on the plane of the sky
of the projected bow-shock tip is $(x_0,y_0)$.}
\label{a1}
\end{figure}

Now, if $(x,y)$ is the observer's reference frame (offsets in
right ascension and declination with respect to a fiducial position), we have
the general situation shown in Figure \ref{a1}. From equation
(\ref{yp}), the shape
of the bow shock on the plane of the sky is
\begin{equation}
y'=Bx'^2\,; \,\,\,{\rm with}\, B=\cos\phi/A^2\,.
\label{xyp}
\end{equation}
Also, the $(x',y')$ and the $(x,y)$ reference
frames obey the standard equations of coordinate transformation
\begin{equation}
x'=(x-x_0)\cos\theta-(y-y_0)\sin\theta\,,
\label{xpxy}
\end{equation}
\begin{equation}
y'=(x-x_0)\sin\theta+(y-y_0)\cos\theta\,,
\label{ypxy}
\end{equation}
where $(x_0,y_0)$ are the coordinates of the head of the
projected bow shock, and $\theta$ is the angle between
the projected symmetry axis and the $y$-axis (see Fig. \ref{a1}).

\subsection{The velocity field}

Let us now consider the motion on the plane of the sky of the post-bow shock
gas. Let us first assume that the gas ahead of the bow shock is at rest.
Following Raga et al. (1997), it is then straightforward to show that,
locally, the gas moves in the direction of the vector normal to the shock.
For a highly radiative shock, because of the large resulting compression
of the material, the cooling region closely follows the motion of the shock.
Therefore, the magnitude of its velocity is:
\begin{equation}
v=v_{bs}\sin\alpha\,,
\label{vvbs}
\end{equation}
where $\alpha$ is obtained from the relation
\begin{equation}
\tan\alpha=\frac{dr}{dw}\,.
\label{alpha}
\end{equation}
The projection of this velocity on the plane of the sky for the points
along the observed edge of the bow shock is
\begin{equation}
v_{x'}=\pm v_{bs}\sin\alpha\cos\alpha\sqrt{1-\tan^2\alpha\tan^2\phi}\,,
\label{vxp}
\end{equation}
\begin{equation}
v_{y'}=-v_{bs}\frac{\sin^2\alpha}{\cos\phi}\,,
\label{vyp}
\end{equation}
while its projection along the line of sight is $v_r=0$, consistent
with the fact that we are seeing the edge of the bow shock, in which
the line of sight is tangent to the bow-shock surface.
The derivation of equations (\ref{vvbs}-\ref{vyp})
is given by Raga et al. (1997; see their eqs. [16] and [17]).

Combining equations  (\ref{yr}),  (\ref{rw}), (\ref{xyp}),
(\ref{alpha}), (\ref{vxp}) and (\ref{vyp})
one finds
\begin{equation}
v_{x'}=v^0_{x'}+\frac{2Bx'}{1+4B^2x'^2}v_{bs}\cos\phi\,,
\label{vxpp}
\end{equation}
\begin{equation}
v_{y'}=v^0_{y'}-\frac{v_{bs}\cos\phi}{1+4B^2{x'}^2}\,,
\label{vypp}
\end{equation}
where we have also introduced $v^0_{x'}$ and $v^0_{y'}$, which
represent the two components of a possible motion (with respect
to the observer) of the environment
within which the bow shock is propagating.

Then, using equations (\ref{xpxy}) and (\ref{ypxy}) with
$x_0=-17.68$mas, $y_0=365.48$mas and $\theta=-9^\circ.7$ (the average
of epochs 1-3 of the parameters given in
Table \ref{tabfits}), we calculate the positions $(x',y')$ from
the maser positions of Table 1. Similarly, using the rotations
\begin{equation}
v_{x'}=v_x\cos\theta-v_y\sin\theta\,,
\label{vxp2}
\end{equation}
\begin{equation}
v_{y'}=v_x\sin\theta+v_y\cos\theta\,,
\label{vyp2}
\end{equation}
we obtain the $v_{x'}$ and $v_{y'}$ motions of the masers (in the
primed reference system, aligned with the bow shock, see figure
\ref{a1}) from the observed proper motions given in Table 1. The
resulting velocities as a function of the $x'$ coordinate are
shown in Figure \ref{a4}. In this figure, we also plot the
total velocity $v_T=\sqrt{v_{x'}^2+v_{y'}^2}$ on the plane
of the sky.

\end{appendix}

\label{lastpage}

\bsp

\end{document}